\documentclass[review]{elsarticle}

\usepackage{moreverb}
\usepackage{amsmath}
\usepackage{amsfonts}
\usepackage{amssymb}
\usepackage{graphicx}
\usepackage{graphics}
\usepackage[citecolor=red]{hyperref}
\usepackage{subfig}
\usepackage{url}
\usepackage[draft]{changes}

\usepackage[figuresright]{rotating}
\usepackage{algpseudocode}

\usepackage{color}
\usepackage{xparse}

\newsavebox{\fminipagebox}
\NewDocumentEnvironment{fminipage}{m O{\fboxsep}}
 {\par\kern#2\noindent\begin{lrbox}{\fminipagebox}
  \begin{minipage}{#1}\ignorespaces}
 {\end{minipage}\end{lrbox}%
  \makebox[#1]{%
    \kern\dimexpr-\fboxsep-\fboxrule\relax
    \fbox{\usebox{\fminipagebox}}%
    \kern\dimexpr-\fboxsep-\fboxrule\relax
  }\par\kern#2
 }

\newcommand\BibTeX{{\rmfamily B\kern-.05em \textsc{i\kern-.025em b}\kern-.08em
T\kern-.1667em\lower.7ex\hbox{E}\kern-.125emX}}

\journal{ArXiV}

\usepackage{natbib}
\usepackage{amssymb}
\usepackage{lineno}
\usepackage{url}
\usepackage{graphicx}
\usepackage{graphics}
\usepackage{framed}
\usepackage{subfig}
\usepackage{longtable}
\usepackage{multirow}
\usepackage{rotating}
    {\pagebreak\global\pdfpageattr\expandafter{\the\pdfpageattr/Rotate 90}}%
    {\pagebreak\global\pdfpageattr\expandafter{\the\pdfpageattr/Rotate 0}}%

\newcommand\bcmdtab{\noindent\bgroup\tabcolsep=0pt%
  \begin{tabular}{@{}p{10pc}@{}p{20pc}@{}}}
\newcommand\ecmdtab{\end{tabular}\egroup}

\begin{document}
\begin{frontmatter}

\title{Revisiting Link Prediction: Evolving Models and Real Data Findings}

\author[add1]{Marcelo Mendoza\corref{cor1}}
\ead{marcelo.mendoza@usm.cl}
\cortext[cor1]{Corresponding author}

\author[add2]{Mat\'ias Estrada}
\ead{matias.estrada@skout.org}

\address[add1]{Universidad T\'ecnica Federico Santa Mar\'ia, Santiago, Chile}

\address[add2]{Skout Inc., Santiago, Chile}

\begin{abstract}
The explosive growth of Web 2.0, which was characterized by the creation of online social networks, has reignited the study of factors that could help us understand the growth and dynamism of these networks. Various generative network models have been proposed, including the Barab\'asi-Albert and Watts-Strogatz models. In this study, we revisit the problem from a perspective that seeks to compare results obtained from these generative models with those from real networks. To this end, we consider the dating network Skout Inc. An analysis is performed on the topological characteristics of the network that could explain the creation of new network links. Afterwards, the results are contrasted with those obtained from the Barab\'asi-Albert and Watts-Strogatz generative models. We conclude that a key factor that could explain the creation of links originates in its cluster structure, where link recommendations are more precise in Watts-Strogatz segmented networks than in Barab\'asi-Albert hierarchical networks. This result reinforces the need to establish more and better network segmentation algorithms that are capable of clustering large networks precisely and efficiently.
\end{abstract}


\end{frontmatter}




\section{Introduction}
\label{intro}

The explosive growth of Web 2.0, which is characterized by its consolidation of online social networks, has brought millions of users into participating in these networks, sharing their life experiences and establishing new relationships through these platforms. The relationships created in these networks can be modeled as large graphs. Their evolution and dynamism represent an interesting challenge for the complex network community; this community is dedicated to proposing and discussing the significance of generative models that could precisely explain the growth of these networks.
	
The most widely discussed generative models are those developed by Barab\'asi-Albert \citep{Barabasi99} and Watts-Strogatz~\citep{watts98}; each of these models has their own strengths and weaknesses in their ability to explain topological characteristics of real networks. Specifically, the Watts-Strogatz model is able to model large but small-diameter networks, which are properties that can be effectively observed in real networks, whereas the Barab\'asi-Albert networks follow an evolving model of preferential attachment with a tendency to generate networks with few vertices that contain a high number of links, where the majority of vertices have low connectivity. This approach has been used as a very effective model for biological networks~\citep{ravasz2002}.

A growing interest in establishing better methods for link predictions is using a microscale approximation of the network evolution problem. This approximation has driven investigations by the complex network community in their efforts to not only find models that effectively fit the dynamic laws of social networks but can also provide the possibility of link recommendation, which is also known as link prediction and is particular useful in online social networks. 

In this study, we perform analyses on link prediction methods based on the topological characteristics of real data from the dating network Skout. Subsequently, we contrast the results obtained with those observed from the Barab\'asi-Albert and Watts-Strogatz generative models. A number of experimental configurations were evaluated, including intra-cluster recommendations in segmented networks. The contribution of this study is the evaluation of the effectiveness of these models in light of the results observed from data of a real network. The problem is analyzed from a purely topological standpoint, which contrasts the classical standpoint of FOAF (friend-of-a-friend) recommendations with those based on segmented networks. 

This article is organized as follows. 
In Section~\ref{sec:relwork} we discuss related work. 
Background is summarized in Section~\ref{back}.
The case study is presented in Section~\ref{case-study}.
Results from synthetic data are presented in Section~\ref{synth} and 
discussed in Section~\ref{disc}.
Finally, we conclude in Section~\ref{conc} with a brief discussion about some open questions and future work. 

\section{Related Work}
\label{sec:relwork}

Link prediction is mainly addressed using transitivity properties of the graph also known as friend-of-a-friend approaches~\citep{silva2010}. 
The rationale of this kind of approaches is the following: 
If a user A is friend of a user B and a user B is friend of a user C, then A and C probably will be friends~\citep{kossinets2006}. 
However, the success of link prediction approaches is limited because real online social networks tend to be very sparse~\citep{delgenio2011}, 
meaning that the total amount of potential links to be created is much greater than the links that are actually created. 
Sparsity turns the link prediction problem into a very dificult problem returning success improvements over a random predictor from 3\% to 54\%~\citep{liben2003}.

A number of different types of features are available for this problem~\citep{lin1998}. 
By exploiting locality we do not require the full graph from being stored. 
One of the simplest locality features is the Common Neighboors index~\citep{newman2001}. 
This index considers the amount of common neighboors that two nodes have. 
When the number of common neighboors is high, then is more likely that those nodes will create a link in the future~\citep{kossinets2006}. 
However, this feature is biased by the number of neighboors that a given node registers.
A more robust feature is the Jaccard index~\citep{jaccard1912} which compares the common neighboors cardinality with 
the cardinality of the union of both neighboors, getting a value that represents the proportion between the cardinality of these two sets. 
Other features used are the Salton index~\citep{salton1983} and the Sorensen index~\citep{sorensen1948} that are mainly used in ecological networks~\citep{linyuan2010}. 
Finally, another well known locality measure is the Adamic \& Adar index~\citep{ada2003}.

Graph-based features considers the whole network to be calculated. 
SimRank index~\citep{jeh2002} is calculated using a random walk process which is propagated through the graph with a decay factor. 
It is an expensive feature because it needs several passes through the whole network.  
Then, unexpensive and conventional graph-based features as degree or transitivity are more recommended for link prediction~\citep{linyuan2010}. 
Finally, the combination of global/local scope is also explored for this task, using approaches as HITS (Hypertext Induced Topic Search) 
which comes from the information retrieval community, proposed by Kleinberg~\citep{kleinberg} for web page ranking. 
An undirected version of HITS can be explored for link prediction purposes in undirected graphs, giving us an authority feature for each node of the graph. 

There are more information sources that are useful for link prediction tasks. 
For instance, by recovering text from user messages, it is possible to create more sophisticated user descriptors~\citep{roth2010}. It is also possible to characterize neighboors by recovering text and then using this information for user description~\citep{zhou2009}.
Transactional registers has been also explored for link prediction~\citep{xlang10}. 
These methods tend to involve high computational costs due to the necessity of maintain content-based indexes or log-repositories. These methods are out of the scope of this study. 

Regarding network segmentation methods, there has been recent interest in the exploration of spectral clustering methods for link prediction. \cite{symeonidis} explored the use of segmented networks for link prediction, proving in synthetic networks that this approach is feasible. The study is close in aim to our article but it does not consider an evaluation regarding evolving graph models. In addition the comparison is constrained only to the top-1 recommended node.

Link prediction has been explored in several domains, showing that is still a relevant task.
For instance, different algorithms has been developed to predict new links in protein networks~\citep{lu2009}, energy grid networks~\citep{murata2008}, medical co-author networks~\citep{plos14}, disease networks~\citep{disease14}, among others.
Further details about this topic can be read in the survey of L\'u~\citep{lu11}. 

\section{Background}
\label{back}

Link prediction is the problem of inferring whether potential edges between pairs of vertices 
in a graph will be present or absent in the near future. 
In a link prediction task, we first assume the existence of an original observed graph, with a completely known set of vertices and a partially known set of edges. 
Accordingly, the link prediction task is to infer the rest of the edges.

Let $G = (V,E)$ be an undirected graph. 
At a given point in time $t_0$, we assume that all vertices $v \in V$ are known but only a subset of $E$ is known. 
Let $E^{obs}$ be the subset of edges $e \in E$ that is known at $t_0$, and let $E^{miss}$ be the subset of unobserved edges at $t_0$, such that $E^{miss} = E \setminus E^{obs}$. 
Link prediction is to predict the edges in $E^{miss}$ from $G = (V,E^{obs})$. 

A natural way to address link prediction is to use a ranking-based strategy. 
A ranking-based strategy considers two steps to rank links. 
For a given vertex $u$, a vertex retrieval step is conducted, 
using a locality constraint to recover a list of proximal vertices to $u$. 
Then, the list is sorted according to a given scoring measure.

A number of graph-based measures can be used for scoring. We explore the perfomance of three of them: Normalized degree, authority, and transitivity. 
These features are query independent, that is to say, these measures define a collection of pointwise estimations at vertex level. 
The value of a measure of this collection is the same for the whole graph. 
The definitions of these measures are given below.

\paragraph{Degree coefficient} Let $\Gamma (u)$ be the neighborhood of a vertex $u$ and let $\mid \Gamma (u) \mid$ be the cardinality of this set, also known as the degree of $u$. 
We define a degree coefficient by normalizing $\mid \Gamma (u) \mid$ with the maximum degree of $G$. Formally:

\[
\tt{Degree}(u) = \frac{\mid \Gamma (u) \mid}{\tt{Max}_{v \in G} \mid \Gamma (v) \mid}.
\]

\paragraph{Transitivity coefficient} Let $\Gamma (u)$ be the neighborhood of a vertex $u$. 
The transitivity coefficient $\tt{Transitivity}(u)$ (a.k.a. clustering coefficient) is the ratio between the number of links in $\Gamma (u)$ and the maximum number of such links. 
If $\Gamma (u)$ has $e_u$ links, we have:

\[
\tt{Transitivity}(u) = \frac{e_u}{\frac{\mid \Gamma (u) \mid \cdot (\mid \Gamma (u) \mid - 1)}{2}} .
\]

\paragraph{Authority} HITS (Hypertext Induced Topic Search) coefficients come from the information retrieval community and were proposed by \cite{kleinberg} for web page ranking. 
The idea is that pages that have many links pointing to them are called authorities and pages that have many outgoing links are called hubs. 
Good hubs point to good authority pages, and vice-versa. 
Lets $\tt{hub}(u)$ and $\tt{auth}(u)$ be hub and authority coefficients for a vertex $u$. 
The following equations can be solved through an iterative algorithm that addresses the fixed point problem defined by:

\[
\tt{hub(u) = \sum_{v \in G \mid u \rightarrow v} \tt{auth}(v)} , 
\]

\[
\tt{auth(u) = \sum_{v \in G \mid v \rightarrow u} \tt{hub}(v)} .
\]

In undirected graphs, both coefficients have the same value. 
In this study, we denote this feature as authority, given that a dating network is an undirected graph.

\section{Link prediction in a real network: the Skout Inc. case}
\label{case-study}

We explore properties of the problem of link prediction in a real network provided directly by Skout Inc.. After recording each link created until January 1st 2014, we generated a network with 3,855,389 links among 1,920,015 active users. We define an active user as one that, after creating their account, has added at least one friend during the period. 

\begin{figure}[h!]
\begin{center}
\includegraphics[width=0.7\columnwidth]{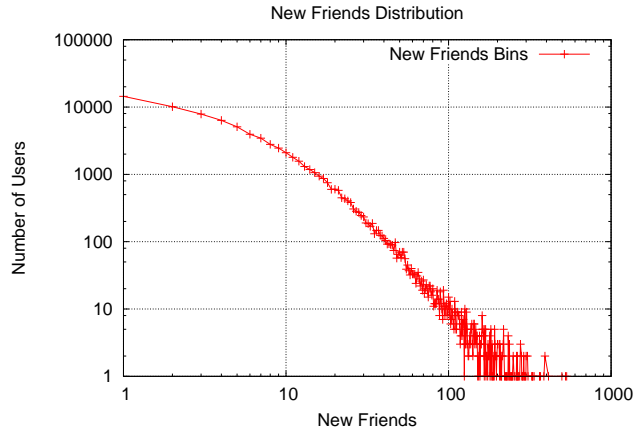}
\end{center}
\caption{Number of links created per active user during the observation period.}
\label{fig-2}
\end{figure}

Afterwards, we recorded links created during the first 25 days of January 2014 between active users. A total of 582,119 new links were recorded solely from users who were active during 2013. A total of 76,848 users added at least one new friend during the observation period. The distribution of new friends per user is shown in Figure~\ref{fig-2}.

As shown in Figure~\ref{fig-2}, the distributions of new friends per user follow the law of “the rich get richer”: only a few users added a high number of new friends; the majority of users added only a few new friends, confirming that link prediction is a very difficult task due to the imbalance between potential friends and actual friends. 

\begin{figure}[h!]
\begin{center}
    \includegraphics[width=0.7\columnwidth]{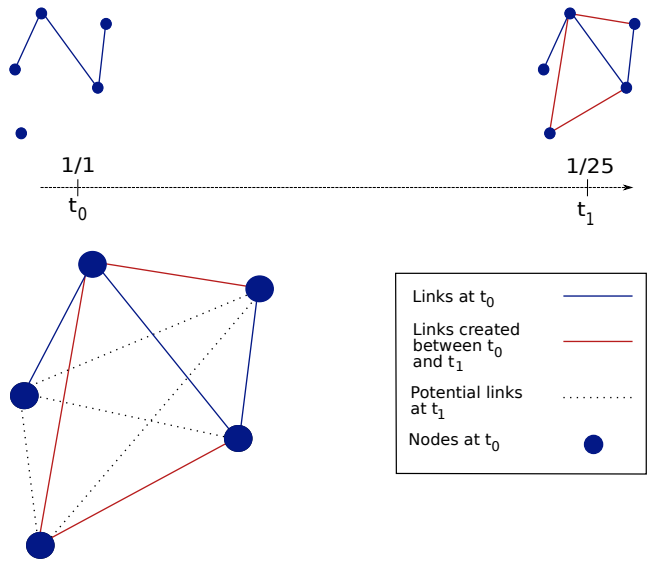}
  \caption{Case study methodology. Blue links and nodes depict the graph at $t_0$. Red links depict links created during the observation period. Links depicted with dashed lines indicates potential unobserved links. In our strategy, dashed links are labeled as false link instances and red links as real link instances.}
  \label{fig-1}
\end{center}
\end{figure}

The collection of links that could have potentially been created during this period corresponds to the group of links that was not created until January 1st 2014 among active users. Then, we analyzed the properties of separability that exist between the group of potential links created versus the group of not-created links; the logic of the methodology used in this study is explained in the scheme shown in Figure~\ref{fig-1}.

Each link in the group of potential links is classified as either real (created) or not (not created), where the authority, degree, and transitivity scores for both the target user and the candidate user are calculated. In this experiment, the target user is the user who is looking for a new friend, and the candidate user is the potential new friend that could be recommended to the target user. 
Table~\ref{tab:inf} shows the results of information gain from separating the two classes.

\begin{table}
\caption{Information Gain values for each feature considered in our data set. Target users features are depicted with a subindex equals to one. Candidate users are depicted with a subindex equals to two.}
\label{tab:inf}
\begin{center}
\begin{minipage}{0.8\textwidth}
\begin{tabular}{lc}
\hline \hline
Feature& \hspace{1cm} Information Gain\\
\hline 
Authority 2& 0.9155\\ 
Authority 1& 0.5367\\
Transitivity 2& 0.1192\\ 
Degree 2& 0.1054\\ 
Transitivity 1& 0.0392\\ 
Degree 1& 0.0227\\
\hline\hline 
\end{tabular}
\end{minipage}
\end{center}
\end{table}

As shown in Table~\ref{tab:inf}, the most relevant characteristic for this problem is the authority of the candidate user. Because the first vertex corresponds to the target user (the one who accepts or declines the new relationship), the authority of the candidate user is a measure of the visibility (and therefore popularity) of the user for the rest of the network. However, the characteristics of the target user are not extremely relevant, which indicates that the present connectivity of the user is not necessarily an indicator of how connected the user will be in the future. We should note that an algorithm that predicts links should be capable of making recommendations to newly target users in the network, i.e., users with few neighbors and a short user history; this challenge is also referred to as a cold start. Because of this, the characteristics of the target user are not as relevant as those of the candidate user. In this strategy, we always consider that it is the user with fewer friends who gets the most recommendations to equalize the size of the network neighborhoods. This focus is paralleled with information retrieval, where the query (the target user) is considered as a document with few descriptors, and therefore, the descriptors of the candidate documents are primarily exploited on being recommended in the list of answers.

\section{Results from synthetic data}
\label{synth}

Networks with 10,000 vertices were generated based on three generative link models: random-based graphs models, also known as Erd\'os-R\'enyi graphs~\citep{erdos60}, Watts-Strogatz~\citep{watts98}, with a rewired probability equal to 0.1, and Barab\'asi-Albert~\citep{Barabasi99}. For each of these models, we performed a pruning process according to a factor taken as an experimental parameter that could control the ratio of links eliminated to all the links in the network. Later, link prediction was performed on the eliminated links, where a recommended and eliminated link was considered a success, and a recommended but kept (not-eliminated) link is considered a failure. In this way, the pruning factor controls the ratio of real solutions to the total candidates. A low pruning factor, which corresponds to a smaller ratio of links to be predicted over the total candidates, makes the problem more difficult.

	Each of the networks was segmented with a directed k-clustering algorithm (k-means for list of arcs), with k values ranging between 3 and 7. Then, using one of the three candidate ranking functions performed on the Skout network (authority, degree, and transitivity), an intra-cluster ranking process was performed. We were thus able to evaluate the impact of network segmentation on the quality of the recommendations.

	Tables 2 and 3 show the average precision and recall results for the first ten recommendations, which are micro-averages, i.e., the results correspond to averages for all the nodes in a network. Table 2 shows the results for a pruning factor of 0.1, and Table 3 shows the results for a pruning factor of 0.25. More configurations were examined in these experiments, which incorporated a greater number of small networks and different pruning factors. The P@N and R@N curves, with N values ranging between 1 and 10 for all configurations, can be seen in greater detail for some small networks (up to 1,000 nodes) at the following url [\url{http://104.236.107.4:17264}]~\footnote{We will replace the ip number by a .org url in the camera ready version}.

\begin{table}
\label{tab:I}
\begin{turn}{90}
\begin{minipage}{\textwidth}
\hspace*{-1cm}
\begin{tabular}{cccccccc}
\hline \hline
& \hspace{0.5mm} k \hspace{1mm}
& Erd\'os-R\'enyi \hspace{1mm}
& Barab\'asi-Albert \hspace{1mm}
& Watts-Strogatz \hspace{1mm}
& Erd\'os-R\'enyi \hspace{1mm}
& Barab\'asi-Albert \hspace{1mm}
& Watts-Strogatz \\
& 
& P@10 & P@10 & P@10& R@10 & R@10& R@10 \\
\hline
\multirow{5}{*}{\rotatebox{90}{Authority}} & 3& 0.0126& 0.0604& 0.0516& 0.0001& 0.0022& 0.0081 \\
 & 4& 0.0121& 0.0462& 0.0798& 0.0005& 0.0015& 0.0169 \\
 & 5& 0.0121& 0.0467& 0.1114& 0.0005& 0.0010& 0.0174 \\
 & 6& 0.0110& 0.0328& 0.1310& 0.0005& 0.0007& 0.0232 \\
 & 7& 0.0104& 0.0424& 0.1552& 0.0008& 0.0010& 0.0298 \\
\hline

\multirow{5}{*}{\rotatebox{90}{Degree}}& 3& 0.0110& 0.0612& 0.0312& 0.0001& 0.0026& 0.0072 \\
 & 4& 0.0110& 0.0464& 0.0513& 0.0004& 0.0012& 0.0143 \\
 & 5& 0.0110& 0.0465& 0.0400& 0.0005& 0.0010& 0.0143 \\
 & 6& 0.0112& 0.0326& 0.0746& 0.0005& 0.0007& 0.0212 \\
 & 7& 0.0112& 0.0412& 0.1066& 0.0008& 0.0010& 0.0268 \\
\hline

\multirow{5}{*}{\rotatebox{90}{Transitivity}}& 3& 0.005­0& 0.0406& 0.0236& 0.0001&0.0026& 0.0074 \\
 & 4& 0.0120& 0.0412& 0.0311& 0.0005& 0.0014& 0.0143 \\
 & 5& 0.0102& 0.0405& 0.0400& 0.0005& 0.0010& 0.0122 \\
 & 6& 0.0126& 0.0333& 0.0441& 0.0005& 0.0007& 0.0204 \\
 & 7& 0.0114& 0.0384& 0.0472& 0.0008& 0.0010& 0.0212 \\
\hline\hline 
\end{tabular}
\begin{center}
\textbf{Table 2}. Precision and recall results with pruning rate = 0.10.
\end{center}
\end{minipage}
\end{turn}
\end{table}

\begin{table}
\label{tab:II}
\begin{turn}{90}
\begin{minipage}{\textwidth}
\hspace*{-1cm}
\begin{tabular}{cccccccc}
\hline \hline
& \hspace{1mm} k \hspace{1mm}
& Erd\'os-R\'enyi \hspace{1mm}
& Barab\'asi-Albert \hspace{1mm}
& Watts-Strogatz \hspace{1mm}
& Erd\'os-R\'enyi \hspace{1mm}
& Barab\'asi-Albert \hspace{1mm}
& Watts-Strogatz \\
& 
& P@10 & P@10 & P@10& R@10 & R@10& R@10 \\
\hline
\multirow{5}{*}{\rotatebox{90}{Authority}} & 3& 0.0272& 0.0712& 0.0988& 0.0018& 0.0033& 0.0144 \\
& 4& 0.0241& 0.0688& 0.1548& 0.0021& 0.0025& 0.0254 \\
& 5& 0.0280& 0.0612& 0.2028& 0.0022& 0.0019& 0.0255 \\
& 6& 0.0282& 0.0575& 0.2616& 0.0014& 0.0018& 0.0339 \\
& 7& 0.0276& 0.0510& 0.2901& 0.0017& 0.0013& 0.0399 \\
\hline 

\multirow{5}{*}{\rotatebox{90}{Degree}}& 3& 0.0264& 0.0791& 0.0744& 0.0018& 0.0033& 0.0122 \\
& 4& 0.0224& 0.0689& 0.1135& 0.0020& 0.0024& 0.0178 \\
& 5& 0.0264& 0.0616& 0.1486& 0.0022& 0.0018& 0.0231 \\
& 6& 0.0270& 0.0579& 0.2235& 0.0014& 0.0018& 0.0336 \\
& 7& 0.0272& 0.0520& 0.2735& 0.0017& 0.0013& 0.0377 \\
\hline 

\multirow{5}{*}{\rotatebox{90}{Transitivity}}& 3& 0.0244& 0.0512& 0.0485& 0.0016&0.0033& 0.0102 \\
& 4& 0.0212& 0.0522& 0.0603& 0.0022& 0.0026& 0.0124 \\
& 5& 0.0270& 0.0614& 0.0850& 0.0022& 0.0019& 0.0146 \\
& 6& 0.0292& 0.0507& 0.1015& 0.0014& 0.0018& 0.0147 \\
& 7& 0.0264& 0.0472& 0.1177& 0.0017& 0.0013& 0.0141 \\
\hline \hline
\end{tabular}
\begin{center}
\textbf{Table 3}. Precision and recall results with pruning rate = 0.25.
\end{center}
\end{minipage}
\end{turn}
\end{table}

\section{Discussion}
\label{disc}

In this study, we addressed the problem of link prediction, which segments networks to later analyze each resulting partition in terms of degree, authority, and transitivity. Several of the results from the experiments using model data demonstrated certain relevant characteristics. One characteristic is related to the link pruning parameter because it is also related to the network evolution time. That is, a low pruning factor emulates a short period of observation and thus, a more difficult problem because worse results would be obtained compared with that of higher pruning factors. This result is due to predictability improving as the limits of observation increase for prediction problems. Specifically for our experiments, increasing the pruning ratio raises the number of possible links that are candidates for selection; each of the candidate links added corresponds to occurrences of links already existing in the original network, and thus, it is expected that the results would improve when they are selected.

The results demonstrate that the process of segmenting the networks is key to link prediction because dissimilar results are obtained from different segmentation routines. Such is the case with the Watts-Strogatz generative model, which achieved a precision of approximately 0.1 with three clusters and approximately 0.3 with seven clusters. The optimum number of clusters depends on the nature of the network, the number of vertices, and the density of the network, among other factors.

Most of the methods of link prediction use locality from a FOAF (friend-of-a-friend) standpoint. However, the results obtained in this research based on transitivity exhibited worse results than those based on authority and degree. It is possible that when using locality, transitivity restricts the predictions to the graph of nearest neighbors, whereas clustering uses global network characteristics, which are only partially approximated by a FOAF standpoint. Barab\'asi-Albert-type networks give worse results from segmentation, where prediction results worsen as the number of clusters increases. Barab\'asi-Albert networks are essentially hierarchical, and a flat clustering segmentation method is therefore not extremely suitable for this type of structure.

The measures used to characterize and select candidates that behave best depend on the network type. In Barab\'asi-Albert-type networks, predictions based on degree are slightly better than those based on authority. The opposite is observed in Watts-Strogatz-type networks; predictions based on authority are somewhat better than those based on degree. In both network types, the worst precision is achieved using transitivity.

When comparing these results to those obtained from Skout Inc., we should remember that in the latter case, the best properties of prediction were shown in the authority, degree, and transitivity scores (in that order). These results are similar to those obtained from a Watts-Strogatz-type generative model, which suggests that this model could approximate the structure of the Skout network, provide a means to perform link prediction on a segmented network, and complement FOAF strategies with ranking strategies based on intra-cluster authority. 

\section{Conclusions}
\label{conc}

This study revisited the problem of link prediction from a perspective that aims to unify results obtained from real networks and those from generative models. We conclude that the recommendations provided from FOAF are overvalued because nearest-neighbor structures are unable to explain the growth and dynamism of a network on their own. Instead, the use of network segmentation methods can explain the creation of links between users that might not be among the nearest-neighbors, which highlights how the dynamism and evolution of a network is explained via a combination of local (nearest neighbor) and global (clusters) factors. These results emphasize the need for more and better network segmentation algorithms capable of working on large data collections. There is also a need to consolidate link prediction models capable of effectively combining local and global factors.

\section*{Acknowledgment}
Marcelo Mendoza was supported by project FONDECYT 11121435 and by FB0821 Centro Cient\'ifico Tecnol\'ogico de Valpara\'iso.

\bibliographystyle{elsarticle-num}

\begin{thebibliography}{00}


\bibitem[\protect\citename{Adamic and Adar, }2003]{ada2003} 
Adamic, L. A., Adar, E. (2003) \emph{Friends and neighbors on the Web}, Social Networks, 25(3), 211-230.


\bibitem[\protect\citename{Barab\'asi and Albert, }1999]{Barabasi99} Barab\'asi, A. L., Albert, R. (1999) \emph{Emergence of scaling in random networks}. Science, 286, 509-512.

\bibitem[\protect\citename{Erd\'os and R\'enyi, }1960]{erdos60} Erd\'os, P., R\'enyi, A. (1960) \emph{On the evolution of random graphs}, Publications of the Mathematical Institute of the Hungarian Academy of Sciences, 5, 17-61.

\bibitem[\protect\citename{Genio \emph{et al.}, }2011]{delgenio2011} Genio, C. I. Del, Gross, T., Bassler, K. E. (2011) \emph{All scale-free networks are sparse}, Physical Review Letters, 107, 178701, 1-4.

\bibitem[\protect\citename{Jaccard, }1912]{jaccard1912} Jaccard, P. (1912) \emph{The distribution of the flora in the alphine zone}, The New Phytologist, XI, 37-50.

\bibitem[\protect\citename{Jeh and Widom, }2002]{jeh2002} Jeh, G., Widom, J. (2002) \emph{SimRank: a measure of structural-context similarity}, In Proceedings of the ACM International Conference on Information and Knowledge Management, KDD, 1-11.

\bibitem[\protect\citename{Kaya and Poyraz, }2014]{disease14} Kaya, B., Poyraz, M. (2014) \emph{Supervised link prediction in symptom networks with evolving case}, Measurement, 56, 231-238.

\bibitem[\protect\citename{Kleinberg, }1998]{kleinberg} Kleinberg, J. (1998) \emph{Authoritative Sources in a Hyperlinked Environment}, in Proceedings of the Symposium on Discrete Algorithms, SODA, 668-677.

\bibitem[\protect\citename{Kleinberg, }2006]{kossinets2006} Kossinets, G. (2006) \emph{Effects of missing data in social networks}, Social Networks, 28(3), 247-268.

\bibitem[\protect\citename{Liben-Nowell and Kleinberg, }2003]{liben2003} Liben-Nowell, D., Kleinberg, J. (2003) \emph{The link prediction problem for social networks}, In Proceedings of the ACM Conference on Information Knowledge and Management, CIKM, 556-559.

\bibitem[\protect\citename{Lin, }1998]{lin1998} Lin, D. (1998) \emph{An Information-Theoretic Definition of Similarity}, in Proceedings of the International Conference on Machine Learning, ICML, 296-304.

\bibitem[\protect\citename{Linyuan, }2010]{linyuan2010} Linyuan, L. (2011) \emph{Link Prediction in Complex Networks : A Survey}. Physica A: Statistical Mechanics and its Applications, 390(6), 1150-1170.

\bibitem[\protect\citename{L\'u \emph{et al.}, }2009]{lu2009} L\'u, L., Jin, C.-H., Zhou, T. (2009) \emph{Similarity index based on local paths for link prediction of complex networks}, Physical Review. E, Statistical, Nonlinear, and Soft Matter Physics, 80, 046122.

\bibitem[\protect\citename{L\'u and Zhou, }2011]{lu11} L\'u, L., Zhou, T. (2011) \emph{Link prediction in complex networks: A survey}, Physica A: Statistical Mechanics and its Applications, 390(6), 1150-1170.





\bibitem[\protect\citename{Murata and Moriyasu, }2008]{murata2008} Murata, T., Moriyasu, S. (2008) \emph{Link Prediction based on Structural Properties of Online Social Networks}, New Generation Computing, 26(3), 245-257.

\bibitem[\protect\citename{Newman, }2001]{newman2001} Newman, M. E. (2001) \emph{Clustering and preferential attachment in growing networks}, Physical Review. E, Statistical, Nonlinear, and Soft Matter Physics, 64, 025102.

\bibitem[\protect\citename{Ravasz \emph{et al.}, }2002]{ravasz2002} Ravasz, E., Somera, A. L., Mongru, D. A., Oltvai, Z. N., Barab\'asi, A. L. (2002) \emph{Hierarchical organization of modularity in metabolic networks}. Science, 297, 1551-1555.

\bibitem[\protect\citename{Roth \emph{et al.}, }2010]{roth2010} Roth, M., Flysher, G., Matias, Y., Leichtberg, A., Merom, R. (2010) Suggesting Friends Using the Implicit Social Graph Categories and Subject Descriptors. In Proceedings of the ACM International Conference on Knowledge Discovery and Data Mining, KDD, 233-242.

\bibitem[\protect\citename{Salton and McGill, }1983]{salton1983} Salton, G., McGill, M. J. (1983) \emph{Introduction to Modern Information Retrieval}, New York (Vol. 22, p. xv, 448 p.).

\bibitem[\protect\citename{Silva \emph{et al.}, }2010]{silva2010} Silva, N. B., Tsang, I.-R., Cavalcanti, G. D. C., Tsang, I.-J. (2010) \emph{A graph-based friend recommendation system using Genetic Algorithm}, In Proceedings of the IEEE Congress on Evolutionary Computation, 1-7. 

\bibitem[\protect\citename{Sorensen, }1948]{sorensen1948} Sorensen, T. (1948) \emph{A method of establishing groups of equal amplitude in plant sociology based on similarity of species content and its application to analyses of the vegetation on Danish commons}, Kongelige Danske Videnskabernes Selskab, 5(4), 1-34.

\bibitem[\protect\citename{Symeonidis \emph{et al.}, }2013]{symeonidis} Symeonidis, P., Iakovidou, N., Mantas, N., Manolopoulos, Y. (2013) \emph{From biological to social networks: Link prediction based on multi-way spectral clustering}, Data \& Knowledge Engineering, 87, 226-242.

\bibitem[\protect\citename{Watts and Strogatz, }1998]{watts98} Watts, D., Strogatz, S. (1998) \emph{Collective dynamics of 'small-world' networks}, Nature, 393, 6684, 440-442.

\bibitem[\protect\citename{Xiang \emph{et al.}, }2010]{xlang10} Xiang, R., Neville, J., Rogati, M. (2010) \emph{Modeling relationship strength in online social networks}, In Proceedings of the ACM World Wide Web Conference, WWW, 981-990.

\bibitem[\protect\citename{Yu \emph{et al.}, }2014]{plos14} Yu Q., Long C., Lv Y., Shao H., He P. (2014) \emph{Predicting Co-Author Relationship in Medical Co-Authorship Networks}, PLoS ONE 9(7): e101214.

\bibitem[\protect\citename{Zhou \emph{et al.}, }2009]{zhou2009} Zhou, T., Lü, L., Zhang, Y.-C. (2009) \emph{Predicting missing links via local information}. The European Physical Journal B, 71(4), 623-630.

\end{thebibliography}

\end{document}